\documentclass[journal]{IEEEtran}
\usepackage{booktabs}
\usepackage{multirow} 
\usepackage{xcolor,soul,framed} 
\usepackage{url}
\usepackage[hidelinks]{hyperref}

\colorlet{shadecolor}{yellow}
\usepackage[pdftex]{graphicx}
\graphicspath{{../pdf/}{../jpeg/}}
\DeclareGraphicsExtensions{.pdf,.jpeg,.png}

\usepackage{algorithm}
\usepackage{algorithmicx}

\usepackage[noend]{algpseudocode}
\usepackage[caption=false,font=footnotesize]{subfig}

\usepackage[cmex10]{amsmath}
\usepackage{array}
\usepackage{mdwmath}
\usepackage{mdwtab}
\usepackage{eqparbox}
\usepackage{booktabs,tabularx,makecell}
\newcolumntype{Y}{>{\raggedright\arraybackslash}X} 
\usepackage{dblfloatfix}  
\usepackage{subcaption} 
\usepackage[caption=false,font=footnotesize]{subfig} 

\usepackage{amssymb}
\allowdisplaybreaks
\usepackage{booktabs,tabularx}
\usepackage{siunitx}
\begin{document}
\bstctlcite{IEEEexample:BSTcontrol}
    \title{Environment-to-Link ISAC with Space-Weather Sensing for Ka-Band LEO Downlinks}
  \author{Houtianfu Wang,~\IEEEmembership{Student Member,~IEEE,}
        Haofan Dong,~\IEEEmembership{Student Member,~IEEE,}
        Hanlin Cai,~\IEEEmembership{Student Member,~IEEE,}
      Ozgur B. Akan,~\IEEEmembership{Fellow,~IEEE}\\

 \thanks{The authors are with Internet of Everything Group, Department of Engineering, University of Cambridge, CB3 0FA Cambridge, UK.}
  \thanks{Ozgur B. Akan is also with the Center for neXt-generation Communications
(CXC), Department of Electrical and Electronics Engineering, Koç University,
34450 Istanbul, Turkey (email:oba21@cam.ac.uk)}
  }


\maketitle

\begin{abstract}
Ka-band low-Earth-orbit (LEO) downlinks can suffer second-scale reliability collapses during flare-driven ionospheric disturbances, where fixed fade margins and reactive adaptive coding and modulation (ACM) are either overly conservative or too slow. This paper presents a GNSS-free, link-internal predictive controller that senses the same downlink via a geometry-free dual-carrier phase observable at 10~Hz: a high-pass filter and template-based onset detector, followed by a four-state nearly-constant-velocity Kalman filter, estimate $\Delta$VTEC and its rate, and a short look-ahead (60~s) yields an endpoint outage probability used as a risk gate to trigger one-step discrete MCS down-switch and pilot-time update with hysteresis. Evaluation uses physics-informed log replay driven by real GOES X-ray flare morphologies under a disjoint-day frozen-calibration protocol, with uncertainty reported via paired moving-block bootstrap. Across stressed 60~s windows, the controller reduces peak BLER by 25--30\% and increases goodput by 0.10--0.15~bps/Hz versus no-adaptation baselines under a unified link-level abstraction. The loop runs in $\mathcal{O}(1)$ per 0.1~s epoch (about 0.042~ms measured), making on-board implementation feasible, and scope and deployment considerations for dispersion-dominated events are discussed.

\end{abstract}

\begin{IEEEkeywords}
LEO NTN, geometry-free dual-carrier phase, Kalman filtering, outage-aware control, MCS/pilot adaptation, space weather.
\end{IEEEkeywords}

%
\IEEEpeerreviewmaketitle


\section{Introduction}

\IEEEPARstart{K}{a}-band low Earth orbit (LEO) downlinks are vulnerable to short-lived yet severe ionospheric disturbances that perturb the total electron content (TEC), causing fast phase and group-delay excursions that can collapse throughput and trigger endpoint outage \cite{Scintillation_Modeling2022,SpaceWeather_Disturbances2020}. In emerging non-terrestrial networks for 5G-Advanced/6G, such links are a key component, but fixed fade margins and static pilot allocations are either too conservative in benign periods or insufficient during sudden disturbances, motivating risk-aware link adaptation compatible with adaptive coding and modulation (ACM) practice \cite{NTN_Survey2023,ISACSurvey2024,ISACOverview2023,Reactive_ACM_WCM2007}. In this work, we introduce a predictive, link-internal controller that senses the same Ka downlink via dual-carrier phase, forecasts the near-term endpoint-outage probability $P_{\mathrm{out}}(t;H)$, and uses it as a risk gate to trigger discrete MCS down-switches and pilot-time updates on the same path without requiring GNSS or other external sensors in the fast loop.

\begin{table*}[!t]
\centering
\caption{Positioning against representative works (same-path, GNSS-free, timescale, control target, complexity, evaluation).}
\label{tab:positioning}
\begingroup
\setlength{\tabcolsep}{4.2pt} 
\renewcommand{\arraystretch}{1.07} 
\small 
\begin{tabular}{@{}lcccccc@{}}
\toprule
Work family & Task & Sensing & Control & Phys@Ch & Out$\rightarrow$Ctrl & NTN \\
\midrule
DFRC beam/waveform      & Beam / TxCov            & Radar (joint)       & Beams / TxCov          & --   & --                         & $\sim$ \\
RIS/STAR-RIS ISAC       & Env.\ reconfig.         & Passive surface     & Env.\ config           & --   & --                         & $\sim$ \\
GNSS-based TEC maps     & Mapping                  & GNSS (L-band)       & None                   & --   & --                         & $\sim$ \\
NTN sched./handover     & Net-level sched.        & None                & Assoc./beam/spec.      & --   & --                         & $\sim$ \\
Reactive ACM (SNR)      & Rate adapt.             & In-band SNR         & MCS                    & --   & --                         & \checkmark \\
This work               & Risk @ horizon $H$       & In-band GF phase    & MCS + pilot (time)     & Iono $\rightarrow$ margin & KF $\rightarrow P_\text{out}$ & \checkmark \\
\bottomrule
\end{tabular}
\par\vspace{0.35ex}
\footnotesize
Note\textemdash\ \checkmark\ = directly addressed; $\sim$ = partially/indirectly addressed or deployability unclear; -- = not addressed. 
GF phase = geometry-free dual-carrier phase on the same Ka downlink; 
Phys@Ch = physics injected at the channel/impairment layer (ionospheric phase $\rightarrow$ required SNR margin with uncertainty penalty); 
Out$\rightarrow$Ctrl = output-to-control coupling (estimated state $\rightarrow$ short-horizon endpoint-outage probability used by the discrete controller); 
Assoc./beam/spec. = association/beam/spectrum; 
TxCov = transmit covariance; 
MCS = modulation and coding scheme.
\endgroup
\end{table*}

Most ISAC/NTN studies emphasize waveform or beam co-design, scheduling, or average-performance optimization, but they rarely close the loop from in-band disturbance sensing to per-link, per-epoch risk control at second scales \cite{DFRC_Outage_TWC,STAR_RIS_ISAC_TWC,NearField_XL_MIMO_TWC,NTN_Survey2023,LEO_Beam_HO_TWC,LEO_HO_DRL2024}. We implement this controller as a low-complexity, link-internal pipeline that (i) senses the Ka-band downlink being adapted, (ii) forecasts near-future endpoint-outage risk, and (iii) actuates discrete modulation-and-coding (MCS) and pilot time to trade throughput against predicted outage under an explicit risk constraint. Here ISAC is link-internal: the same Ka downlink being adapted is sensed via a geometry-free dual-carrier phase; a short-horizon ($\le\!60$\,s) endpoint-outage probability is produced; consistent with the sub-minute F-region TEC response to flare EUV/X-ray forcing and with avoiding seconds-scale scintillation gating~\cite{OHare2025Atmosphere,Vani2019TVT}; and discrete MCS/pilot updates are gated at 10\,Hz with $\mathcal{O}(1)$ runtime. This fast loop remains in-band and same-path; see Sec.~II–IV for contrasts and rationale \cite{ISACOverview2023,Reactive_ACM_WCM2007}. National Oceanic and Atmospheric Administration (NOAA)’s Geostationary Operational Environmental Satellites (GOES) X-ray Sensor (XRS) products provide slow-time priors or drive evaluation traces only; the fast loop is GNSS-free and onboard-feasible.

In-band sensing uses a geometry-free dual-carrier Ka-band phase observable sampled at 10\,Hz. In practice, the two tones can be implemented as narrowband pilots or dedicated beacons on the same downlink. Ka-band propagation/beacon campaigns demonstrate stable same-path phase tracking for dual-frequency measurements. A parameterized unit-energy flare template with block-maximum thresholding provides matched-filter detection with controlled false alarm. A nearly-constant-velocity Kalman filter tracks $\Delta\mathrm{TEC}$ and its rate; a short-horizon propagator maps the filtered state to an endpoint-outage probability on the same Ka downlink, specializing reliability-aware ISAC ideas to flare-driven, seconds-scale risk control in this setting \cite{DFRC_Outage_TWC}. A single BLER–margin mapping, learned once and reused, supports a discrete MCS ladder with hysteresis and sets pilot time fractions in an ACM-compatible manner \cite{Reactive_ACM_WCM2007}. This yields lower BLER and higher time-averaged goodput while keeping seconds-scale outage under control. 

We summarize our contributions in three concrete points: 
\begin{itemize}
\item \textbf{Predictive environment-to-link mapping for Ka-band LEO downlinks.} Starting from the same-path dual-carrier geometry-free phase on the Ka downlink, we track $\Delta\mathrm{VTEC}$ and its rate, map them via a physics-informed margin model to the required SNR margin with uncertainty, and output a calibrated, short-horizon endpoint-outage probability $P_{\text{out}}(t;H)$ on that specific link, which directly gates discrete MCS and pilot-time decisions without external sensors or beacons (GNSS-free).

\item \textbf{Lightweight seconds-scale loop for on-board use.}
We design a fully in-band, link-internal ISAC loop with per-cycle cost $O(1)$, running comfortably faster than real time on a single CPU thread, while remaining compliant with existing NR/NTN control primitives (discrete MCS and pilot budgets).

\item \textbf{Physics-informed, event-driven evaluation.} Using real flare-driven morphologies with geometry/SNR slices, we report paired moving-block-bootstrap confidence intervals, ablations on sensing/front-end parameters (e.g., the high-pass constant), and ensemble robustness across elevations and $C/N_0$. We benchmark primarily against no-adapt and tuned reactive-ACM baselines.

\end{itemize}

At a high level, we maximize time-averaged goodput subject to an endpoint-outage constraint via discrete MCS/pilot actuation. Across disturbance amplitudes, elevations, and $C/N_0$ offsets, results show consistent time-averaged goodput gains and lower BLER, with larger instantaneous improvements in stressed intervals. Statistical significance and robustness are established by paired moving-block bootstrap and ablations, and the implementation runs in $\mathcal{O}(1)$ time and memory per step, leaving ample real-time margin for onboard use.

Section~II reviews related work. Section~III presents the system and signal model and states the control problem. Section~IV details sensing, filtering, risk forecasting, and the discrete MCS/pilot controller. Section~V reports detection, estimation, and closed-loop performance with ablations and robustness. Section~VI concludes.

\section{Related Work}

We distinguish estimation/mapping (instantaneous, latency well below $T_c$) from prediction (look-ahead beyond a fraction of $T_c$), and separate network-level ISAC from link-internal, same-path sensing in NTN/LEO. In parallel, semantic communication architectures for the Internet of Space have been proposed at higher protocol layers, using task-oriented or semantic metrics to coordinate heterogeneous space networks rather than per-link physical-layer outage control \cite{Semantic_IoS}.  

In contrast to works that optimize network averages, rely on GNSS-aided probes, or react via SNR-only ACM, our module is a seconds-scale, GNSS-free, same-path predictor that outputs endpoint-outage probability to gate discrete MCS and pilot budgets with $\mathcal{O}(1)$ per-epoch complexity. Table~\ref{tab:positioning} positions this design against representative families along same-path sensing, GNSS independence, time scale, control target, complexity, and evaluation modality, and highlights its focus on ionosphere-aware environment-to-link control for Ka-band LEO downlinks.

\subsection{Integrated Sensing and Communications (ISAC)}
Recent ISAC research has established rigorous frameworks for jointly optimizing communication utility and sensing performance, mostly in terrestrial settings with block-fading or slowly varying channels \cite{ISACSurvey2024,ISACOverview2023}. Representative strands include: (i) DFRC beamforming/waveform co-design with reliability constraints \cite{DFRC_Outage_TWC,DFRC_Waveform2022,DFRC_Robust2019}; (ii) RIS/STAR-RIS assisted ISAC coordinating active–passive beams \cite{STAR_RIS_ISAC_TWC,IRS_ISAC_TWC,STAR_RIS_Secure2024}; and (iii) near-field/XL-MIMO localization/sensing with wideband arrays \cite{NearField_XL_MIMO_TWC,XL_MIMO_Localization2024}. These works formalize multi-objective tradeoffs and propose robust designs under CSI uncertainty. ISAC concepts have also been extended to low-Earth-orbit terahertz inter-satellite channels, where cooperative sensing and communications are analyzed and fundamental performance limits are characterized under hardware impairments, pointing errors, and network interference \cite{CoopISAC_LEO_THz}.

DFRC beamforming and waveform co-design typically actuate spatial covariances/beams and rely on frame-level optimizers (e.g., SDR/SOCP/iterative convexification) \cite{DFRC_Outage_TWC,DFRC_Waveform2022,DFRC_Robust2019}. In contrast, flare-driven ionospheric disturbances at Ka-band induce colored, seconds-scale phase excursions on the very downlink to be protected; the actuation here is discrete MCS/pilot selection with temporal risk gating on a fixed link, and the loop must run at 10~Hz with $\mathcal{O}(1)$ per-epoch complexity.

RIS/STAR-RIS assisted ISAC exposes rich spatial reconfigurability \cite{STAR_RIS_ISAC_TWC,IRS_ISAC_TWC,STAR_RIS_Secure2024}, yet deployable and controllable reflect/refract surfaces along a space-to-ground Ka path are not readily available. We therefore keep the joint-optimization mindset but actuate what is natively available—discrete MCS and pilot time—rather than environment reconfiguration.

Near-field/XL-MIMO ISAC leverages wavefront curvature and beam squint for high-resolution localization/sensing \cite{NearField_XL_MIMO_TWC,XL_MIMO_Localization2024}. These advances target geometric identifiability and array control, whereas the present focus is per-link, seconds-scale reliability under ionospheric disturbances with a fixed antenna interface.


\begin{figure*}[!t]
  \centering
  \includegraphics[width=\textwidth]{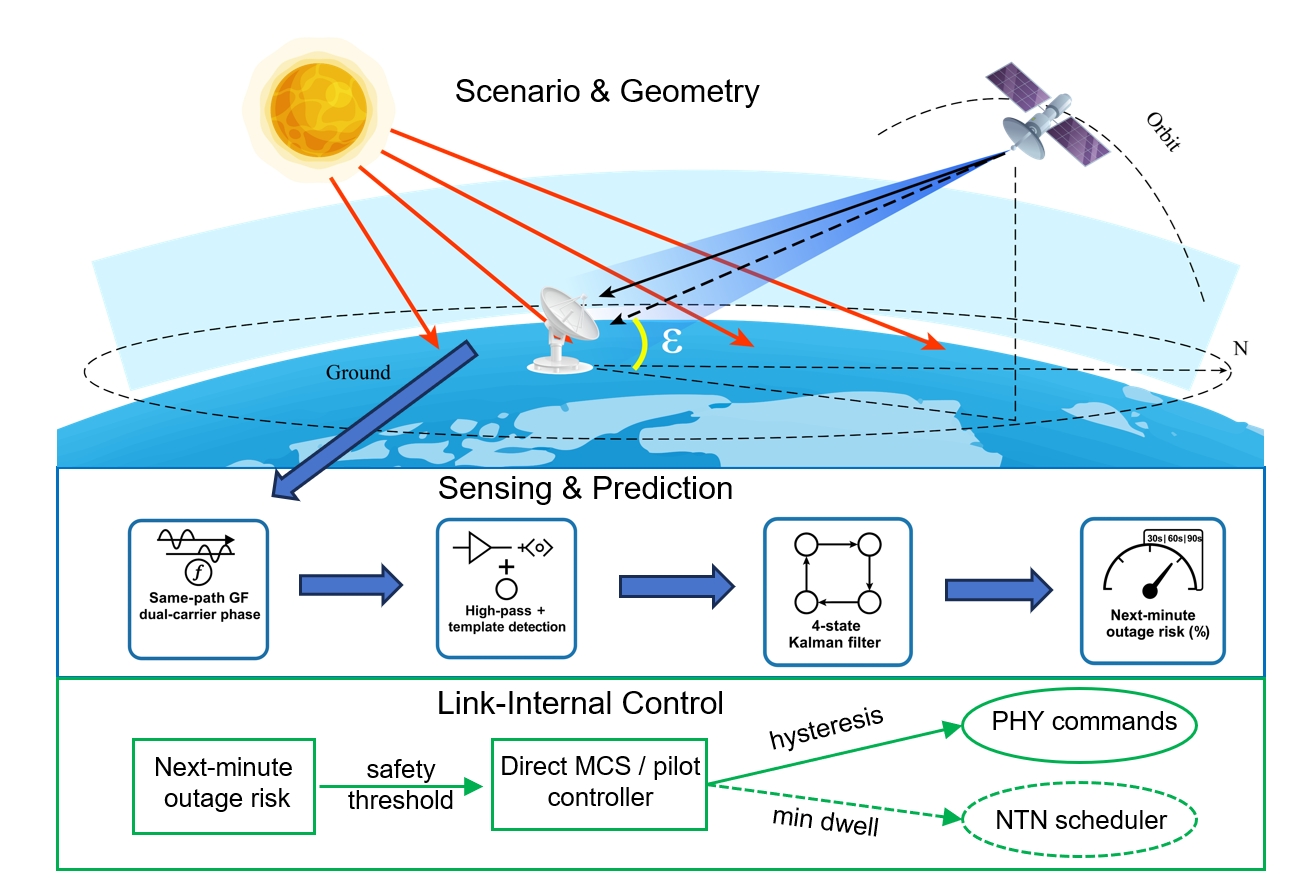}
  \caption{Link-internal ISAC overview (Ka-band LEO). Top—geometry; Middle—same-path GF dual-carrier sensing and risk; Bottom—risk-gated MCS/pilot controller with optional NR-NTN interface.}
  \label{fig:system-overview}
\end{figure*}

\subsection{NTN/LEO Communications: Beam Management, Handover, and Scheduling}
LEO NTN studies cover beam management, frequent handover mitigation, spectrum sharing, and long-horizon scheduling/association under rapid topology changes \cite{NTN_Survey2023,LEO_Beam_HO_TWC,LEO_Scheduling2024}. Typical objectives emphasize time-averaged throughput, handover frequency, or coverage, often with learning-based controllers \cite{LEO_HO_DRL2024,NTN_Learning2023,LEO_Beam_HO_TWC,Reactive_ACM_WCM2007}. Two gaps remain for our target: (i) network-level controllers optimize on tens-of-seconds to minutes and seldom incorporate in-band, same-path observables that flag endpoint outage within a short horizon $H$; and (ii) their actuation concerns association/beam/spectrum rather than per-epoch, risk-gated MCS/pilot decisions at millisecond-level runtime. 

Our module is complementary: it runs locally at seconds scale, exposes a calibrated endpoint-outage probability and low-overhead set-points (MCS/pilots) to protect stressed intervals, and can be consumed upstream as constraints/weights by schedulers and handover logic (e.g., protect windows, rate caps, conservative flags) without replacing existing NTN controllers \cite{Reactive_ACM_WCM2007,LEO_Beam_HO_TWC}.

\subsection{GNSS-Based TEC Sensing and Ionospheric Disturbances}
Dual-frequency GNSS provides mature pipelines for slant/vertical TEC estimation and disturbance mapping, including geometry-free combinations, differential-code-bias calibration, code smoothing, and regional/global products. Popular indices (e.g., ROTI/AATR) and both model-based and data-driven reconstructions are widely used \cite{GNSS_TEC_Mapping2019,GNSS_TEC_Processing2021,Scintillation_Modeling2022}. Space-weather studies further indicate that flare-driven disturbances can emerge rapidly and non-stationarily \cite{SpaceWeather_Disturbances2020,SGD_2023}.

For a Ka-band LEO downlink, a link-internal approach is motivated by objective/geometry mismatch and by latency/coupling considerations. GNSS products target accurate TEC at L-band over diverse geometries, whereas the controller here must protect a specific Ka downlink with same-path observables directly coupled to the impairment and executable at 10~Hz. Many GNSS products operate at slower cadences and are not natively coupled to per-epoch physical-layer actuation (MCS/pilots). We therefore treat GNSS as an optional slow-timescale prior—e.g., to initialize state-space priors or set dynamic risk thresholds—while keeping the fast loop in-band and in-geometry to avoid external dependencies and timing mismatch. In our implementation, GNSS-derived TEC products and GOES X-ray Sensor (XRS) time series are used only offline for calibrating model parameters and for constructing evaluation scenarios; the runtime controller itself relies solely on in-band, same-path dual-carrier phase and does not query these external sensors. This view aligns with recent perspectives that leverage astrophysical messengers as side information for engineered links while keeping fast control loops in-band \cite{11132318}.


\section{System and Signal Model}\label{sec:system}

We consider a Ka-band low-Earth-orbit (LEO) downlink with two closely spaced carriers at $f_1 = 20.2$\,GHz and $f_2 = 19.7$\,GHz. These frequencies lie at the edges of the ITU Ka downlink allocation (19.7--20.2\,GHz) widely used by FSS/HTS systems, including GEO and NGSO deployments (e.g., KA-SAT, OneWeb) \cite{KA_SAT_Satsig, FCC_2025}.
Solar-flare–induced ionospheric disturbances perturb the total electron content (TEC), introducing a dispersive phase shift on the received carriers.
A geometry-free (GF) phase combination isolates this ionospheric signature for sensing and subsequent link adaptation. The default sampling period is $\Delta t = 0.1$\,s. Fig.~\ref{fig:system-overview} summarizes the three-layer loop used throughout: (i) scenario and geometry of the same Ka-band downlink; (ii) same-path sensing and risk estimation from the GNSS-free GF dual-carrier phase at 10 Hz through high-pass + template detection and a 4-state Kalman filter to a next-minute outage risk; and (iii) link-internal control where this risk gates a discrete MCS/pilot controller that issues PHY-level commands, with an optional NR-NTN scheduler interface.


\subsection*{Notation}
$\mathrm{TECU}=10^{16}$\,e$^-$/m$^2$; $c$ is the speed of light; $K_{\rm iono}=40.3$.
Bold letters denote vectors/matrices and $(\cdot)^\top$ denotes transpose.
We use the vertical increment $\Delta\mathrm{VTEC}(t)$ throughout; figures may abbreviate it as $\Delta\mathrm{TEC}$ for brevity. Slant/vertical mapping is handled explicitly below.

\subsection{Propagation and Observation Model}

Under the thin-shell model with Earth radius $R_{\rm E}$, shell height $h_{\rm I}$, and elevation
angle $\varepsilon$, the slant-to-vertical mapping is
\begin{equation}
\label{eq:mapping}
\begin{aligned}
M(\varepsilon) &= \frac{1}{\sqrt{1-\big(\tfrac{R_{\rm E}\cos\varepsilon}{R_{\rm E}+h_{\rm I}}\big)^2}}, \\
\Delta\mathrm{STEC}(t) &= M(\varepsilon)\,\Delta\mathrm{VTEC}(t).
\end{aligned}
\end{equation}
Let $\phi_i(t)$ denote the carrier phase at $f_i$ in radians. The geometry-free (GF) phase observable removes
geometry-dependent and common hardware terms:
\begin{equation}
y(t) \;=\; \phi_1(t) - \phi_2(t) \;=\; k_{\rm GF}\,\Delta\mathrm{STEC}(t) + b(t) + v(t),
\end{equation}
where $b(t)$ is a slowly varying aggregate bias (residual hardware and scintillation leakage) and
$v(t)$ is differential receiver phase noise. The ionospheric coefficient (radians/TECU) is
\begin{equation}
k_{\rm GF} \;=\; -\frac{2\pi K_{\rm iono}\cdot 10^{16}}{c}
\Big(\tfrac{1}{f_1}-\tfrac{1}{f_2}\Big),
\end{equation}
and with $f_1{=}20.2$\,GHz, $f_2{=}19.7$\,GHz we have $k_{\rm GF}\approx 0.0106$\,rad/TECU.
Defining the effective gain $K_{\rm eff}\triangleq k_{\rm GF}M(\varepsilon)$ gives the compact form
\begin{equation}
y(t) \;=\; K_{\rm eff}\,\Delta\mathrm{VTEC}(t) + b(t) + v(t).
\end{equation}

\subsubsection{Noise variance and mapping sensitivity}
Using a standard time–bandwidth argument for carrier/phase tracking, the per-carrier phase variance satisfies
$\sigma_\phi^2 \approx {B_{\rm eff}}/{(C/N_0)}$ with $B_{\rm eff}\!\approx\!{1}/{(2\Delta t)}$ for a first-order estimator.
Hence for the GF difference $v(t)=n_1(t)-n_2(t)$,
\begin{equation}
\mathrm{Var}[v(t)] \approx 2\sigma_\phi^2 \approx \frac{1}{(C/N_0)\,\Delta t},
\end{equation}
i.e., $R \approx 1/\big[(C/N_0)\Delta t\big]$ with $C/N_0$ in linear units.
For uncertainty propagation through the slant–to–vertical mapping, let $a \triangleq R_{\rm E}/(R_{\rm E}+h_{\rm I})$.
Then $M(\varepsilon)=\big(1-a^2\cos^2\varepsilon\big)^{-1/2}$ and
\begin{equation}
\frac{\partial M}{\partial \varepsilon}
= -\,\frac{a^2\,\sin\varepsilon\,\cos\varepsilon}
{\big(1-a^2\cos^2\varepsilon\big)^{3/2}}\,,
\end{equation}
which is dimensionless and useful for linear error propagation.

\subsubsection{Assumptions}
(A1) $\varepsilon \ge 30^\circ$ and elevation rate $<0.5^\circ$/min so $M(\varepsilon)$ is quasi-static over 200\,s windows;
(A2) moderate scintillation is absorbed by $b(t)$; (A3) carrier-phase thermal noise follows
$\sigma_\phi^2 \approx [\,2(C/N_0)\Delta t\,]^{-1}$ per carrier, thus $\mathrm{Var}[v]\approx 1/((C/N_0)\Delta t)$~\cite{9100468};
(A4) phase continuity is maintained; (A5) a cycle-slip monitor enforces $|\Delta(\phi_1{-}\phi_2)|<\pi/2$ per sample.
For example, with $C/N_0{=}52$ dB-Hz (linear $1.58{\times}10^5$) and $\Delta t{=}0.1$ s,
$\mathrm{Var}[v]\!\approx\!6.3{\times}10^{-5}$ rad$^2$ (standard deviation $\approx 8{\times}10^{-3}$ rad),
consistent with the adopted tracking regime.

The state-space model and noise covariances consistent with this observable are given in Sec.~\ref{sec:ssm}.

\subsection{Problem Formulation}
Let the control at epoch $t$ be $u(t)\!=\![\,r(t),\,\eta(t)]^\top\!\in\!\mathcal U$, where
\[
\mathcal{U}\triangleq\{\,r\!\in\!\mathcal{R},\ \eta\!\in\![\eta_{\min},\eta_{\max}]\,\}.
\]
Goodput $T(t)$ is given by \eqref{eq:goodput}, and $m_{\rm avail}^{\rm extra}(t)$, $m_{\rm eff}(t)$ by \eqref{eq:mavail}–\eqref{eq:meff}.
We maximize time–average goodput with a short-horizon outage constraint:
\begin{equation}
\max_{\{u(t)\}\subseteq \mathcal{U}} \ \frac{1}{T}\sum_{t=1}^{T} T(t)
\ \ \text{s.t.}\ \ P_{\rm out}(t{+}H)\le \delta,\ \forall t,
\label{eq:prob-constrained}
\end{equation}
or its Lagrangian surrogate
\begin{equation}
\max_{\{u(t)\}\subseteq \mathcal{U}} \ \frac{1}{T}\sum_{t=1}^{T}
\big[T(t)-\lambda\,P_{\rm out}(t{+}H)\big],\quad \lambda\ge 0.
\label{eq:prob-lagrangian}
\end{equation}

\subsubsection{BLER target $\Rightarrow$ margin threshold}
For target BLER $\beta$ (default $0.10$), the logistic map \eqref{eq:bler_logit} with frozen $(k,m_{\mathrm{piv}})$ implies
\begin{equation}
m_\beta \triangleq m_{\mathrm{piv}} + \frac{1}{k}\,\ln\!\frac{\beta-\mathrm{bler}_{\min}}{\mathrm{bler}_{\max}-\beta}.
\label{eq:mbeta}
\end{equation}

\subsubsection{Per-epoch surrogates (on-board)}
\begin{equation}
\max_{u(t)\in\mathcal{U}}\ \mathbb{E}[T(t)]
\ \ \text{s.t.}\ \ m_{\rm eff}(t)\ \ge\ m_\beta ,
\label{eq:opt}
\end{equation}
\begin{equation}
\max_{u(t)\in\mathcal{U}}\ \mathbb{E}[T(t)]
\ \ \text{s.t.}\ \ \Pr\{m_{\rm eff}(t{+}H)\ge m_\beta\}\ \ge\ 1-\delta,
\label{eq:opt-risk}
\end{equation}
where $P_{\rm out}(t;H)$ in \eqref{eq:pout} (Sec.~\ref{sec:sensing-risk}) gives the equivalent constraint $P_{\rm out}(t;H)\le\delta$; we set $H{=}60$\,s to capture the most hazardous first-minute margin surge (sub-minute F-region response) while smoothing seconds-scale scintillation and keeping scheduler/dwell practical~\cite{OHare2025Atmosphere,Vani2019TVT}.

\subsection{PHY Abstraction and Performance Metrics}
\label{sec:phy-abstraction}

We model the downlink PHY via a small discrete MCS ladder and a single BLER--margin
anchor reused by all policies. The scheduled spectral efficiency is
$r(t)\in\mathcal{R}=\{4.0,\,5.0,\,6.4\}$\,bps/Hz, corresponding to representative
MCS levels (MCS-3: 16QAM-3/4; MCS-4: 64QAM-2/3; MCS-5: 64QAM-5/6). Let
$r_{\max}\triangleq \max\mathcal{R}$ and let $k_r$ denote the rate--margin slope
with units bps/Hz/dB.

\subsubsection*{MCS ladder and BLER anchor}
We score all methods using a single pre-aligned logistic BLER–margin proxy that approximates the selected NR link-level curves. Its parameters (k, mpiv) are learned once from a no-adapt calibration log and then kept fixed for every policy to avoid circular tuning.
Table~\ref{tab:mcs} lists the reproducible ladder: modulation, coding rate,
scheduled $r$, and the logistic parameters $(k,m_{\mathrm{piv}})$.

\begin{table}[htbp]
\caption{Abstracted MCS ladder and BLER anchor (logistic proxy pre-aligned to standard curves; used in all policies).}
\label{tab:mcs}
\centering
\footnotesize
\setlength{\tabcolsep}{4pt}
\renewcommand{\arraystretch}{1.05}
\begin{tabular}{@{} l l l l l l @{}}
\toprule
Idx & Mod. & Code rate & $r$ (bps/Hz) & $k$ & $m_{\mathrm{piv}}$ (dB) \\
\midrule
MCS-3 & 16QAM & $3/4$ & $4.0$ & $\approx 1.01$ & $\approx 0.72$ \\
MCS-4 & 64QAM & $2/3$ & $5.0$ & $\approx 1.01$ & $\approx 0.72$ \\
MCS-5 & 64QAM & $5/6$ & $6.4$ & $\approx 1.01$ & $\approx 0.72$ \\
\bottomrule
\end{tabular}
\end{table}

\subsubsection*{Available and effective margin}
Selecting a lower-rate MCS provides headroom expressed as the available extra margin
\begin{equation}
m_{\rm avail}^{\rm extra}(t)\;=\;\frac{r_{\max}-r(t)}{k_r}.
\label{eq:mavail}
\end{equation}
Given the required margin $m_{\rm req}(t)$ defined by the sensing branch, the effective margin is
\begin{equation}
m_{\rm eff}(t)\;=\;m_{\rm avail}^{\rm extra}(t)-m_{\rm req}(t).
\label{eq:meff}
\end{equation}

\subsubsection*{BLER vs.\ margin (logistic)}
Let $\mathrm{bler}_{\min},\mathrm{bler}_{\max}\in[0,1]$ denote the floor/ceiling.
With slope $k>0$ and pivot (midpoint) $m_{\mathrm{piv}}$ (dB), the BLER mapping is
\begin{multline}
\label{eq:bler_logit}
\mathrm{BLER}\big(m_{\rm eff}\big)\;=\;
\mathrm{bler}_{\min} \\
+\big(\mathrm{bler}_{\max}-\mathrm{bler}_{\min}\big)
\Big(1+\exp\{-k\,(m_{\rm eff}-m_{\mathrm{piv}})\}\Big)^{-1}.
\end{multline}
The pair $(k,m_{\mathrm{piv}})$ is learned once from the no-adapt run and then frozen
for all comparisons to avoid circular tuning.

\subsubsection*{Goodput accounting}
With pilot time fraction $\eta(t)\in[0,1]$, the goodput is
\begin{equation}
T(t) \;=\; \big(1-\eta(t)\big)\,\big(1-\mathrm{BLER}(m_{\rm eff}(t))\big)\,r(t).
\label{eq:goodput}
\end{equation}
We disable online pilot-power adaptation and keep $\alpha(t)\equiv\alpha_{0}$ fixed from the calibration log; consequently $\alpha$ is absorbed into the BLER–margin calibration once and does not vary during evaluation or scoring. In all closed-loop comparisons (Fig.~\ref{fig:bler_thr}), we report the goodput $T(t)$ to account for pilot-time overhead.

\subsubsection*{Justification of the rate--margin slope $k_r$}
With implementation efficiency $\eta_e\!\approx\!0.75$,
$r(\gamma)=\eta\log_2(1+\gamma)$. Linearizing at an operating SINR
$\gamma_0$ (linear) gives
\begin{equation}
\frac{\partial r}{\partial \gamma_{\rm dB}}\Big|_{\gamma_0}
=\eta\,\frac{\ln 10}{10\,\ln 2}\cdot\frac{\gamma_0}{1+\gamma_0}
\quad\text{[bps/Hz/dB]}.
\label{eq:dr_dB}
\end{equation}
Numerically, with $\eta_e{=}0.75$ and $\gamma_0\!\in\![10,30]$ (linear),
$\partial r/\partial\gamma_{\rm dB}\approx 0.226$–$0.241$ bps/Hz/dB,
consistent with the adopted $k_r{=}0.20$\,bps/Hz/dB.
\vspace{0.5ex}

This abstraction links MCS scheduling decisions to achievable throughput under
ionospheric perturbations while remaining compatible with the margin estimates
developed earlier.

\subsection{State-Space Representation}\label{sec:ssm}

For joint tracking of the disturbance and a slowly varying bias, we adopt a four-state constant-velocity model
\begin{equation}
\begin{aligned}
\mathbf x_k &=
\begin{bmatrix}
\Delta\mathrm{VTEC}_k\\ \dot{\Delta\mathrm{VTEC}}_k\\ b_k\\ \dot b_k
\end{bmatrix}, \\
\mathbf x_{k+1} &= \mathbf F\,\mathbf x_k + \mathbf w_k, \\
y_k &= \mathbf h^\top \mathbf x_k + v_k,
\end{aligned}
\end{equation}
with transition matrix and observation vector
\begin{equation}
\mathbf F \;=\; \begin{bmatrix}
1&\Delta t&0&0\\[0.1ex]
0&1&0&0\\[0.1ex]
0&0&1&\Delta t\\[0.1ex]
0&0&0&1
\end{bmatrix},\qquad
\mathbf h \;=\; \begin{bmatrix} K_{\rm eff} & 0 & 1 & 0 \end{bmatrix}^{\!\top}.
\end{equation}
Here $y_k$ denotes the pre-HPF measurement used by the estimator; the HPF is part of the detection branch only, and any associated noise gain is accounted for in the detection calibration. The process noise is
$\mathbf w_k\!\sim\!\mathcal N(\mathbf 0,\,\mathbf Q\,\Delta t)$ with $\mathbf Q\!=\!\mathrm{diag}(q_1,q_2,q_3,q_4)$, and the measurement noise is $v_k\!\sim\!\mathcal N(0,R)$ with $R$ consistent with Assumption~A3. 

\subsubsection*{Process-noise magnitudes}
The adopted diagonal $\mathbf Q=\mathrm{diag}(q_1,q_2,q_3,q_4)$ is consistent with
quiet-to-disturbed ionospheric variations (TEC drifts $\sim$0.1–5 TECU/min) 
and slow hardware bias wander; the values in Table~\ref{tab:defaults} were
validated by NIS (Normalized Innovation Squared) $\approx 1$ and PCRB-tight confidence bands.

\subsection{Default Parameters}

All default constants are consolidated in Table~\ref{tab:defaults} and referenced by later sections.

\begin{table}[htbp]
\caption{Default System and Algorithm Parameters}
\label{tab:defaults}
\centering
\footnotesize
\setlength{\tabcolsep}{3pt}      
\renewcommand{\arraystretch}{1.05} 
\begin{tabularx}{\columnwidth}{@{} l l Y @{}}
\toprule
\textbf{Parameter} & \textbf{Value} & \textbf{Description} \\
\midrule
$\Delta t$        & $0.1$ s                       & Sampling period \\
$\tau_{\rm HP}$   & $200$ s                        & HP time constant \\
$C/N_0$           & $52$ dB-Hz                     & Phase-noise ref. (A3) \\
$q_1$             & $(7{\times}10^{-5})^2$ TECU$^2$/s    & Process noise (TEC) \\
$q_2$             & $(2{\times}10^{-4})^2$ (TECU/s)$^2$/s & Process noise (TEC rate) \\
$q_3$             & $(2{\times}10^{-7})^2$ rad$^2$/s     & Process noise (bias) \\
$q_4$             & $(5{\times}10^{-8})^2$ (rad/s)$^2$/s  & Process noise (bias rate) \\
$w_s$             & $200$ s                        & Template window length \\
$T_{\mathrm{cal}}$   & $4800$ s   &Calibration log length (no-event) \\
$N_b$               & $24$       &Number of non-overlapping blocks for block-max \\
$m_0$             & $3.2$ dB                       & Baseline in \eqref{eq:mreq} \\
$k_1$             & $1.2$ dB/TECU                  & TEC sensitivity \\
$k_2$             & $22$ dB/(TECU/s)               & Rate-of-change sensitivity \\
$r_{\max}$        & $6.4$ bps/Hz                   & Max spectral efficiency \\
$r_{\min}$        & $4.0$ bps/Hz                   & Min spectral efficiency \\
$k_r$             & $0.20$ bps/Hz/dB     & Rate–margin slope in \eqref{eq:rate} \\
$H$               & $60$ s                         & Outage horizon \\
$\eta_{\min,\max}$& $[0.15,\,0.30]$                & Pilot time fractions \\
$m_{\rm sat}$     & $3.0$ dB                       & Saturation in \eqref{eq:eta} \\
$z_{\rm hi}$      & $2.685$                        & Upper trigger; block–max on calibration log; per-window $\alpha$ \\
$z_{\rm lo}$      & $z_{\rm hi}{-}0.7$             & Clear threshold (hysteresis) \\

$T_{\rm dwell,min}$ & $10$ s                       & Minimum state dwell (debounce) \\
$\alpha_{\rm FA}$         & $10^{-3}$                      & Per-window false alarm \\
$\Delta t_{\text{XRS}}$ & 60 s & XRS$\rightarrow\Delta$VTEC mapping step \\
$\chi$ & -- & Solar zenith angle at receiver (rad) \\
$\Delta F_X(t)$ & -- & Detrended GOES XRS-B (1--8\,\AA) irradiance (W\,m$^{-2}$) \\
\bottomrule
\end{tabularx}
\end{table}


\section{Sensing, Risk Forecasting, and Discrete Control}\label{sec:method}
This section details the estimation, detection, forecasting, and discrete control pipeline that
implements the system model of Section~\ref{sec:system}. All symbols and
defaults are as defined in Section~\ref{sec:system} (Table~\ref{tab:defaults}).

GF dual-carrier phase is preferred over raw SNR because it captures dispersive ionospheric dynamics while avoiding AGC/scheduler confounders. A light high-pass with a causal template isolates seconds-scale rise–decay morphologies at constant time. A 4-state constant-velocity Kalman filter (KF) provides uncertainty-aware extrapolation on short windows and runs deterministically on OBCs. A 60\,s look-ahead aligns with the sub-minute F-region response while avoiding seconds-scale scintillation gating~\cite{OHare2025Atmosphere,Vani2019TVT}. A one-step MCS downshift with pilot saturation, stabilized by hysteresis and minimum dwell, minimizes flapping and protects decoder convergence under bursty risk.

\subsection{Calibration and Evidence Protocol}\label{sec:evidence-scope}
We adopt a real-event–driven, physics-informed emulation to preserve external validity while avoiding circular tuning:
\begin{enumerate}
\item Public GOES flare time series drive the disturbance.

\item A causal mapping from GOES XRS (1 min) to $\Delta$VTEC and then to slant-path phase: minute XRS is detrended, a driven–relaxation response yields $\Delta$VTEC (discretized at 60 s: $\Delta\mathrm{VTEC}_k = e^{-\Delta t_{\text{XRS}}/\tau_d}\,\Delta\mathrm{VTEC}_{k-1}
+ \alpha_0 \cos^{\gamma}\!\chi_k \sum_{i\le k} e^{-(t_k - t_i)/\tau_d}\,\Delta F_X(t_i)\,\Delta t_{\text{XRS}}$), thin-shell M($\epsilon$) gives $\Delta$STEC, and the geometry-free phase is formed; the causal high-pass is used only in detection, and the Kalman filter consumes the pre-HPF GF observable. Parameters are calibrated once on disjoint flare days with collocated GNSS TEC and then frozen for all evaluations.

\item The communication layer is scored with a pre-aligned logistic BLER–margin proxy for the selected MCS ladder.

\item All thresholds and curve parameters are locked on a calibration no-event log of length 
$T_{\mathrm{cal}} = N_b \, w_s$ using $N_b$ non-overlapping length-$w_s$ blocks, and remain fixed during evaluation. 
In our runs, $w_s=200$\,s, $N_b=24$, so $T_{\mathrm{cal}}=4800$\,s. No statistic is tuned on the evaluation window.

\end{enumerate}
This protocol fixes strategy-side choices before evaluation and reduces the risk of leakage between calibration and testing. All signals (including the pilot fraction $\eta(t)$) are interpolated onto the same discrete timeline used by the mapping so that BLER and goodput are computed on a shared time base for all policies. \emph{We reuse the single logistic BLER–margin anchor from Sec.~\ref{sec:phy-abstraction}; $(k,m_{\mathrm{piv}})$ are fixed from a no-adapt run and never refit.}

The 10\,Hz loop reads the dual-carrier phase once per slot, computes a decision within the same slot, and applies it to the next slot. The only knobs exposed to the existing scheduler are the rate index $r\in\mathcal{R}$ (the existing MCS indices) and the pilot/training share $\eta(t)$. No new CSI types or feedback messages are introduced; we reuse the current path that already carries rate changes. The latency budget $\Delta$ simply means “sensing readout $\rightarrow$ compute $\rightarrow$ apply at the next slot,” and the horizon $H$ is the look-ahead window before the next scheduling point; neither alters numerology.

\subsection{Ionospheric Disturbance Sensing and Risk Forecasting}\label{sec:sensing-risk}\phantomsection\label{sec:meas-detect}\label{sec:outage}

We fuse three ingredients into a single sensing–forecasting chain with a shared time base and frozen calibration (Sec.~\ref{sec:evidence-scope}): (i) causal matched filtering for event detection, (ii) a four-state KF for state tracking with NIS consistency, and (iii) a Gaussian short-horizon predictor that maps the state and covariance to endpoint outage probability.

\subsubsection{Measurement formation, preprocessing, and detection}
We form the GF observable with the effective gain $K_{\text{eff}}$, apply the causal high-pass filter, and construct a causal zero-mean unit-energy parameterized template $\tilde g$ that is fixed \emph{a priori} and independent of evaluation data. For no-event density estimation we discard the first $L{-}1$ causal-prefix samples in each length-$L$ block before forming the histogram of $z_{\text{norm}}$ to avoid an artificial pile-up at zero. We then compute the matched-filter output and normalize it using the pre-event standard deviation so that $z_{\text{norm}}[k]=z[k]/\sigma_{\text{eff}}$. To control per-window false alarms under colored noise, we use block-maximum thresholding with $L=\lfloor w_s/\Delta t\rfloor$. The first post-$t_0$ threshold crossing of $z_{\rm norm}$ defines the time-to-first-alarm (TTFA). When the number of blocks is small, an extreme-value (Gumbel) fit to the block maxima can extrapolate the operational quantile.

\subsubsection{State estimation and consistency}
We estimate $\mathbf x_k=[\,\Delta\mathrm{VTEC}_k,\ \dot{\Delta\mathrm{VTEC}}_k,\ b_k,\ \dot b_k\,]^\top$
with a constant-velocity KF and diagonal $\mathbf Q$; the measurement variance is $R=2\sigma_\phi^2$ per Assumption~(A3).
Consistency is monitored via the normalized innovation squared
\begin{equation}
\mathrm{NIS}_k \;\triangleq\; \frac{\nu_k^2}{S_k},
\label{eq:nis}
\end{equation}
with
\begin{equation}
\nu_k \;\triangleq\; y_k \;-\; \mathbf h^\top \,\widehat{\mathbf x}_{k|k-1},
\label{eq:nu}
\end{equation}
\begin{equation}
S_k \;\triangleq\; \mathbf h^\top \,\mathbf P_{k|k-1}\,\mathbf h \;+\; R .
\label{eq:S}
\end{equation}
For a correctly tuned linear–Gaussian filter with scalar observation, $\mathrm{NIS}_k$ is $\chi^2(1)$–distributed in expectation; the KF posterior covariance $\mathbf P_k$ coincides with the Bayesian posterior Cramér–Rao bound (PCRB), so $1.96\,\sqrt{[\mathbf P_k]_{1,1}}$ provides a PCRB–tight $95\%$ band for $\Delta\mathrm{VTEC}$.

\subsubsection{Short-horizon outage prediction}
Let $\boldsymbol\xi(t)\!\triangleq\![\,\Delta\mathrm{VTEC}(t),\,\dot{\Delta\mathrm{VTEC}}(t)\,]^\top$.
Propagate its mean/covariance to $t{+}H$ to obtain
$\boldsymbol\mu_\xi$ and $\mathbf P_{\xi\xi}$, with $\mu_1{=}\boldsymbol\mu_\xi[1]$, $\mu_2{=}\boldsymbol\mu_\xi[2]$.
Linearizing \eqref{eq:mreq} at $\boldsymbol\mu_\xi$,
\begin{align}
\mu_{\rm req}
&\approx m_0 + k_1[\mu_1]^+ + k_2|\mu_2| + \Delta m_\phi\!\big(\mathbf P_{k+H}\big), \nonumber\\
\sigma_{\rm req}^2
&\approx \mathbf g^\top \mathbf P_{\xi\xi}\,\mathbf g,\qquad
\mathbf g \triangleq
\begin{bmatrix}
k_1\,H(\mu_1)\\[0.2ex]
k_2\,\operatorname{sgn}(\mu_2)
\end{bmatrix}.
\label{eq:mreq_mean_var}
\end{align}
The offset $m_0$ and slopes $k_1$ and $k_2$ are chosen once based on a pre-event baseline interval in the same trace, so that $\mu_{\mathrm{req}}$ increases monotonically with both the incremental TEC and its rate of change. These coefficients are then kept fixed for all subsequent experiments without re-tuning.

Comparing with the instantaneous available margin
$m_{\rm avail}^{\rm extra}(t)=(r_{\max}-r(t))/k_r$ from \eqref{eq:mavail},
the endpoint outage probability at look-ahead $H$ is
\begin{equation}
P_{\rm out}(t;H) \;=\;
\Phi\!\left(\frac{\mu_{\rm req}-m_{\rm avail}^{\rm extra}(t)-c}{\sigma_{\rm req}}\right),
\label{eq:pout}
\end{equation}
where $\Phi(\cdot)$ is the standard normal CDF and $c$ is a one-time baseline constant. Here, $\mathcal{T}_{\text{pre}}$ denotes a contiguous pre-event baseline interval in the same trace, taken over $[t_0 - T_{\text{pre}}, t_0)$ after an initial warm-up. The constant $c$ is determined numerically so that the empirical mean of $P_{\text{out}}(t;H)$ over $\mathcal{T}_{\text{pre}}$ equals $0.10$, which we interpret as a light-risk baseline. Once calibrated on this pre-event segment, $c$ is kept frozen for all subsequent times in that trace.

A window-level probability over $[t,t{+}H]$ can be approximated from finely sampled endpoints via a Dunn–\v{S}id\'ak rule:
\begin{equation}
\label{eq:pwin}
P_{\rm win}(t;H) \;\approx\;
1-\prod_{i=1}^{n}\Big(1-P_{\rm out}(t_i;H)\Big),
\end{equation}
assuming weak dependence among the sampled endpoints $\{t_i\}\subset[t,t{+}H]$.
When $P_{\rm out}(t_i;H)$ is nearly constant across the grid, \eqref{eq:pwin} reduces to $P_{\rm win}\approx 1-(1-\bar p)^n$. By default we report the endpoint probability $P_{\rm out}$ to avoid double counting; $P_{\rm win}$ is only used when a window-level risk is explicitly needed.

\subsection{Risk-Aware Discrete Control Policy}\label{sec:policy}
Given the KF output $(\widehat{\mathbf x}_k,\mathbf P_k)$ with
$\widehat{\mathbf x}_k=[\,\widehat{\Delta\mathrm{VTEC}}_k,\widehat{\dot{\Delta\mathrm{VTEC}}}_k,\widehat b_k,\widehat{\dot b}_k\,]^\top$,
we compute a sensing-driven required margin (dB)
\begin{equation}
m_{\rm req}(t)
= m_0
  + k_1\,[\widehat{\Delta\mathrm{VTEC}}(t)]^{+}
  + k_2\,\big|\widehat{\dot{\Delta\mathrm{VTEC}}}(t)\big|
  + \Delta m_\phi(t),
\label{eq:mreq}
\end{equation}
where $[x]^+\!\triangleq\!\max\{x,0\}$ and the uncertainty penalty is
\begin{equation}
\Delta m_\phi(t)
= 10\log_{10}\!\big(1+\kappa_\phi\,K_{\rm eff}^2\,[\mathbf P_k]_{1,1}\big),
\label{eq:dmphi}
\end{equation}
with $\kappa_\phi\triangleq \rho\,\gamma_{0,\mathrm{lin}}$ (default $\rho{=}1$;
$\gamma_{0,\mathrm{lin}}{=}10^{\gamma_{0,\mathrm{dB}}/10}$). Once calibrated on a
pre-event segment, $\rho$ is fixed and does not depend on $C/N_0$ (dB-Hz).

The scheduled spectral efficiency follows a clipped linear law
\begin{equation}
r(t)=\big[r_{\max}-k_r\,m_{\rm req}(t)\big]_{[r_{\min},\,r_{\max}]},
\label{eq:rate}
\end{equation}
then a discrete MCS state machine with hysteresis and minimum dwell.
Pilot time fractions follow saturating laws, while pilot power is fixed at $\alpha_{0}$
\begin{align}
\eta(t)&=\eta_{\min}
+\frac{m_{\rm req}(t)}{m_{\rm req}(t)+m_{\rm sat}}\big(\eta_{\max}-\eta_{\min}\big),
\label{eq:eta}\\
\alpha(t) \equiv \alpha_{0}.
\label{eq:alpha}
\end{align}
The effective margin entering the BLER map is $m_{\rm eff}(t)=m_{\rm avail}^{\rm extra}(t)-m_{\rm req}(t)$
(cf.~\eqref{eq:mavail}--\eqref{eq:meff}); goodput is given by \eqref{eq:goodput}. The online controller that implements \eqref{eq:mreq}--\eqref{eq:alpha} together with the outage-gating rule \eqref{eq:pout} is summarized in Alg.~\ref{alg:isac}.

\begin{algorithm}[htbp]
  \caption{Online risk gating and discrete control.}
  \label{alg:isac}
  \footnotesize
  \begin{algorithmic}[1]
    \Require KF prior $(x_0,P_0)$; thresholds $(T_\alpha,z_{\mathrm{hi}},z_{\mathrm{lo}})$; gate $\tau_{\mathrm{gate}}$; dwell $T_{\mathrm{dwell}}$; rate ladder $\{r_{\min},r_{\max}\}$
    \Ensure Control $(\eta,\alpha)$ and rate $r$
    \State $y \gets \phi_1 - \phi_2$; \textsc{KF-PredictUpdate} $\Rightarrow (\hat{x}, P)$
    \State $z_{\mathrm{norm}} \gets \mathrm{MF}(\mathrm{HPF}[y])/\hat{\sigma}_{\mathrm{pre}}$; \texttt{detected} $\gets$ rising cross of $T_\alpha$
    \State $m_{\mathrm{eff}} \gets$ \eqref{eq:meff}; \quad $P_{\mathrm{out}}(t;H) \gets$ \eqref{eq:pout}
    \If{\texttt{detected} \textbf{and} $P_{\mathrm{out}} > \tau_{\mathrm{gate}}$}
       \State \textsc{Adapt}: soften via \eqref{eq:bler_logit} at $m_{\mathrm{eff}}{+}\delta m$
       \State dwell on $z_{\mathrm{norm}}$ to set $r \in \{r_{\min}, r_{\max}\}$
    \Else
       \State $r \gets r_{\max}$
    \EndIf
    \State Set $(\eta,\alpha) \gets$ \eqref{eq:eta}–\eqref{eq:alpha}; compute BLER/goodput via \eqref{eq:bler_logit}, \eqref{eq:goodput}
    \State \textit{Note:} 10\,Hz, per-tick complexity $\mathcal{O}(1)$.
  \end{algorithmic}
\end{algorithm}

\subsubsection{Policy instantiation and baselines}
Time-series predictors (e.g., ARIMA or small RNNs) can serve as one-step margin forecasters. To represent this route without model-selection confounds, we include a \textit{prediction-only margin gate} as a dedicated baseline. Our predictive policies are:
(i) \textbf{adapt-1}: keep $r(t)$ fixed (MCS-4) but soften the BLER mapping after detection via a one-time right shift $\delta m>0$ of the logistic input, modeling modest coding/ARQ gain without re-fitting $(k,m_{\text{piv}})$;
(ii) \textbf{adapt-1+2}: in addition, trigger a single down-switch to MCS-3 when the high-threshold condition is met with positive slope, with hysteresis ($z_{\rm lo}$) and minimum dwell times to prevent chattering. Thresholds and dwell times follow Table~\ref{tab:defaults}.

For context, we compare against operator-style baselines:
(1) \textbf{Reactive-average}: short moving-average SNR thresholds for up/down rate changes; $\eta(t)$ fixed.
(2) \textbf{Fixed-safety}: reserve a constant safety margin; $\eta(t)$ fixed.
(3) \textbf{Reactive ACM (SNR-threshold)}: use the smoothed margin as a gate with hysteresis and dwell (enter $=0$\,dB, exit $=+1$\,dB, minimum hold $=45$\,s); eligibility is over the full timeline (no external time windowing).
(4) \textbf{Prediction-only margin gate (mean forecast)}: propagate the Eq.~(25) surrogate to obtain the mean required margin $\mu_{\rm req}(t{+}H)=m_0+k_1 z_H+k_2 s_H$ (with $z_H\!\approx\!\mathrm{MA}(z_{\rm norm})+\dot z\,H$ and $s_H\!\approx\!\mathrm{MA}(|\Delta\phi_{\rm GF,HP}|)$), define $\mu_{\rm eff}(t{+}H)=m_{\rm extra}^{\rm avail}(t)-\mu_{\rm req}(t{+}H)$, and gate on $\mu_{\rm eff}(t{+}H)\ge m_\beta$ using the same hysteresis and minimum dwell as above.

All methods share the same scoring pipeline (BLER--margin map), dwell/hysteresis settings, and calibration; the baselines use the same fixed hyperparameters as the proposed policies. The first three baselines are purely reactive, whereas the fourth uses a mean forecast but does not invoke the probabilistic map $P_{\text{out}}$.

\subsubsection{Fairness protocol}
In the main comparison we lock the pilot time to a constant $\eta(t)\equiv\eta_{0}$ for all methods (proposed and baselines),
eliminating extra degrees of freedom beyond the discrete MCS state machine. As a complementary $\eta$-shared ablation, all methods are also evaluated under the same $\eta(t)$ trajectory computed by \eqref{eq:eta} from the common $m_{\rm req}(t)$, so pilot time has identical freedom across methods.

\subsubsection{Solution complexity}
Because the BLER map is non-convex and $r$ is discrete, we solve via the heuristic laws \eqref{eq:rate}–\eqref{eq:alpha} with $\mathcal{O}(1)$ work per epoch (4-state KF + closed-form risk), which yields statistically significant goodput gains (Sec.~\ref{sec:results}, bootstrap $p<10^{-3}$). If \eqref{eq:rate} is infeasible even at $r{=}r_{\min}$ and $\eta{=}\eta_{\max}$, we saturate to that corner and minimize the margin shortfall.

\section{Experiments and Results}\label{sec:results}

This section evaluates the end-to-end ISAC pipeline developed in Sections~\ref{sec:system}--\ref{sec:method}. 
Unless otherwise stated, parameters follow Table~\ref{tab:defaults}: two Ka-band carriers ($f_1{=}20.2$\,GHz, 
$f_2{=}19.7$\,GHz), sampling $\Delta t{=}0.1$\,s, elevation $\varepsilon{=}40^\circ$, $C/N_0{=}52$\,dB-Hz, 
and nominal $\mathrm{VTEC}_0{=}12$\,TECU. Nuisance statistics (phase-noise variance, block maxima, logistic BLER anchor) are learned on a preceding no-event calibration log and then frozen to avoid circular tuning.

Unless stated otherwise, blue denotes the baseline (no-adapt), orange denotes adapt-1, and green denotes adapt-1+2. Thresholds and reference lines are red dashed. The KF 95\% band coincides with the PCRB; when overlaid, the PCRB curve is purple.

\begin{figure}[htbp]
  \centering
  \subfloat[Matched-filter output with block-max threshold ($\alpha_{\rm FA}=10^{-3}$).]{%
    \includegraphics[width=\columnwidth]{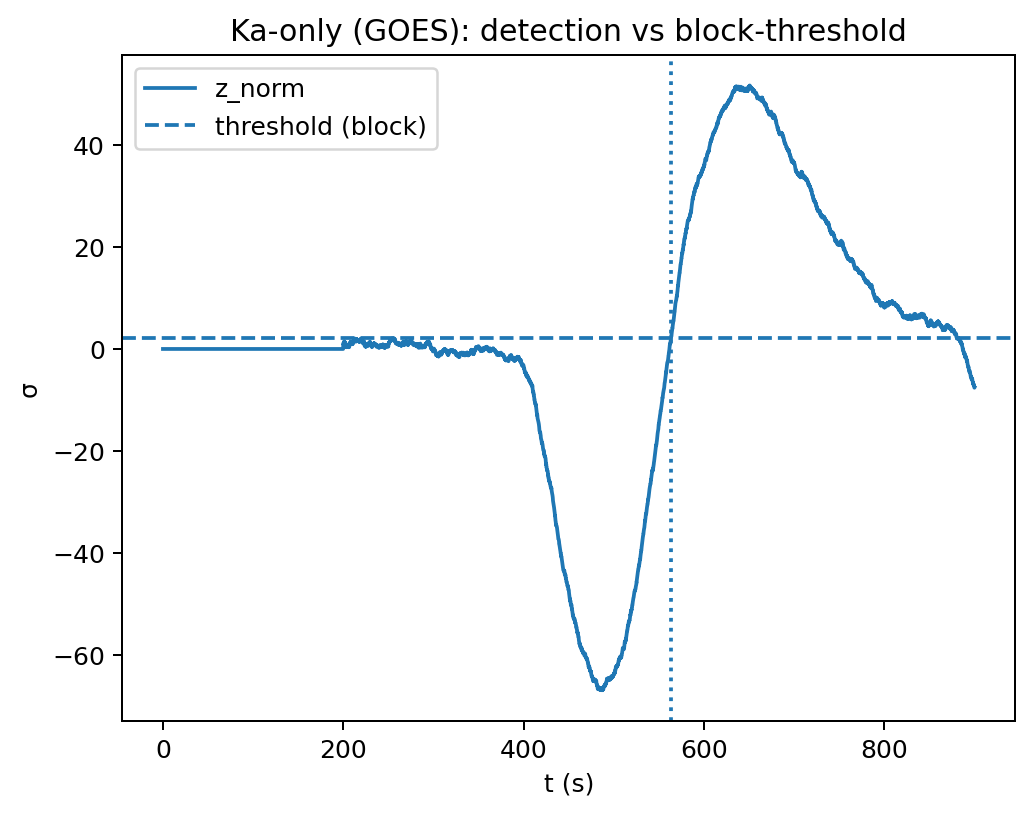}%
    \label{fig:det}}\\[-0.3ex]
  \subfloat[$\Delta$VTEC KF estimate with 95\% band and Bayesian PCRB (at $C/N_0{=}52$ dB-Hz, $\varepsilon{=}40^\circ$).]{%
    \includegraphics[width=\columnwidth]{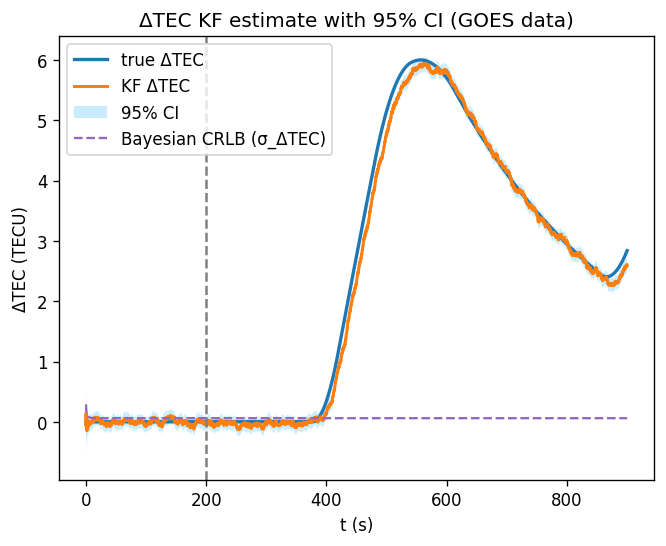}%
    \label{fig:kf_pcrb}}
  \caption{Detection and sensing quality. (a) Operating trace with block-max threshold; (b) KF posterior band is numerically PCRB-tight, validating the uncertainty passed to risk gating.}
  \label{fig:detect_cal}
\end{figure}

\subsection{Reporting Views and Statistical Methods}\label{sec:eval}
We follow the evaluation protocol of Section~\ref{sec:evidence-scope}. Public GOES XRS-B (1--8\,\AA) drives the disturbance as a causal morphology driver (not a TEC nowcast); parameters are pre-calibrated and frozen, which reduces the influence of moderate mapping error on comparative results under the shared scoring and thresholds. Two official feeds are used throughout: the National Centers for Environmental Information (NCEI) one-minute archive and the Space Weather Prediction Center (SWPC) operational one-minute feed. In both cases we take the long-wavelength B band (1–8 Å, 0.1–0.8 nm) irradiance, keep the native one-minute averages in W/m$^{2}$, align timestamps in UTC, and drop duplicate minutes only; no smoothing or resampling is applied. All detection thresholds, filter settings, and the BLER–margin parameters are fixed by a pre-event calibration and reused unchanged in all experiments.

We pre-register the following analysis plan. Primary outcomes are full-timeline averages over the entire evaluation window without any gating: time-averaged goodput and BLER. Secondary outcomes are risk-gated analyses using pre-registered outage
gates $\tau_{\text{gate}}\in\{0.2,0.3,0.4\}$ applied to $P_{\text{out}}(t;H)$. We use paired moving-block bootstrap (block length fixed by variance-stabilizing heuristics; $B{=}2000$) and report 95\% percentile CIs and effect sizes (Cohen's $d$). To control family-wise error across the three gates, we apply the Holm--Bonferroni procedure: if $p_{(1)}\le p_{(2)}\le p_{(3)}$
are the uncorrected $p$-values, the corrected thresholds are $\alpha_{\mathrm{sig}}/3,\alpha_{\mathrm{sig}}/2,\alpha_{\mathrm{sig}}$ for $p_{(1)},p_{(2)},p_{(3)}$, respectively. Both uncorrected and corrected $p$-values are reported. The gates reflect operational SLAs where endpoint outage beyond 0.2--0.4 is actionable.

\subsection{Setup and Detector Calibration}

As calibrated once in Sec.~\ref{sec:evidence-scope}, we use a no-event calibration log with $T_{\mathrm{cal}}=N_b w_s$, and freeze all thresholds thereafter. Fig.~\ref{fig:detect_cal}(a)–(b) visualize the operating trace with the block-max threshold in (a) and the KF-based state estimate with a PCRB-tight 95\% band in (b).

Figs.~\ref{fig:detect_cal}(a)–(b) jointly tie the detector's operating point to the declared false-alarm budget and validate the propagated uncertainty: $z_{\rm norm}$ stays near zero before $t_0$, then crosses the empirical threshold $T_\alpha$ on the rising edge after $t_0$—the time-to-first-alarm (TTFA) depends on event strength and SNR—while the KF 95\% band widens through the crest and coincides with the Bayesian PCRB. We use a single calibrated trigger, i.e., $T_{\alpha}\triangleq z_{\mathrm{hi}}$ (Table~\ref{tab:defaults}).

The high-pass constant is fixed to $\tau_{\mathrm{HP}}=200\,\mathrm{s}$.
The sweep $\{300,600,900,1200\}\,\mathrm{s}$ in Table~\ref{tab:ablation_tauhp} pertains to $\tau_{\mathrm{HP}}$. All primary results fix $\tau_{\mathrm{HP}}=200\,\mathrm{s}$; Table~V shows a broad plateau; deployments prioritizing faster gate timing may set $\tau_{\mathrm{HP}}=300$–$600\,\mathrm{s}$ without changing the ordering.

\subsection{Sensing Quality and Calibration}

\subsubsection{Sensing Performance: KF Estimate and PCRB}
State estimation uses the four-state constant-velocity KF of Section~\ref{sec:system} on the pre-HPF 
observable $y_k$, with $R$ from the $C/N_0$ law and diagonal $\mathbf Q$ from Table~\ref{tab:defaults}. 
Under linear–Gaussian assumptions, the KF posterior covariance $\mathbf P_k$ equals the Bayesian posterior PCRB, yielding a PCRB-tight $95\%$ band for $\Delta\mathrm{VTEC}$ of width $1.96\sqrt{[\mathbf P_k]_{1,1}}$.

Fig.~\ref{fig:kf_pcrb} illustrates PCRB-tightness at $C/N_0{=}52$\,dB-Hz and $\varepsilon{=}40^\circ$, supporting the reliability of the uncertainty propagated to the communication side.

\subsubsection{PHY Calibration: BLER vs Effective Margin}
\begin{figure}[htbp]
  \centering
  \includegraphics[width=\columnwidth]{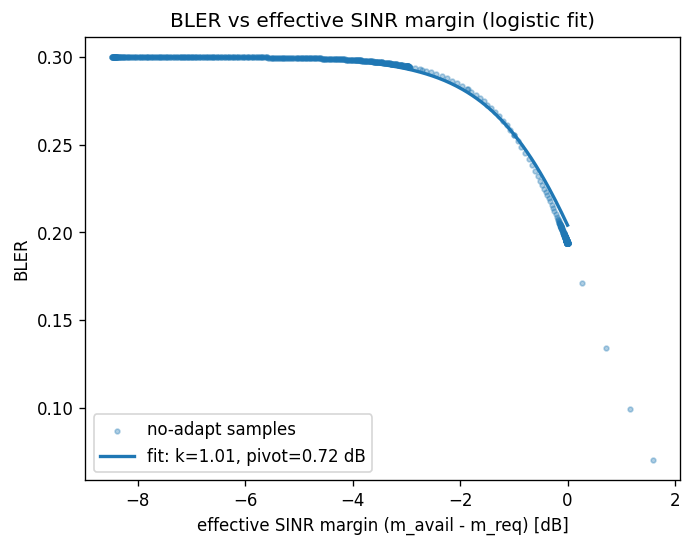}
  \caption{BLER vs effective SINR margin (calibration logistic fit).}
  \label{fig:blerfit}
\end{figure}
The PHY abstraction is anchored by a single logistic fit of BLER versus effective margin $m_{\rm eff}=m_{\rm avail}^{\rm extra}-m_{\rm req}$ using the no-adapt baseline (fixed MCS-4, $r=5.0$\,bps/Hz). Only the slope and pivot are learned (e.g., $k{\approx}1.01$, $m_{\mathrm{piv}}{\approx}0.72$\,dB) and frozen for all policies. The empirical scatter and the single logistic fit are shown in Fig.~\ref{fig:blerfit}; this one-time calibration is reused verbatim by all policies. Varying $(k,m_{\mathrm{piv}})$ within the 95\% bootstrap CI changes gated $\Delta T$ by $\le 0.004$\,bps/Hz and $\Delta$BLER by $<0.001$, suggesting that the aggregate metrics are not strongly driven by the exact anchor fit within this range.

\subsection{Closed Loop Performance}

\subsubsection{Short-Horizon Reliability: Endpoint Outage Probability}

Given horizon $H{=}60$\,s, we evaluate the endpoint outage via \eqref{eq:pout} by propagating $(\widehat{\mathbf x}_k,\mathbf P_k)$ to $t{+}H$ and linearizing the margin mapping to obtain $(\mu_{\rm req},\sigma^2_{\rm req})$.


\begin{figure*}[htbp]
  \centering
  \subfloat[2017\textendash09, $P_{\mathrm{out}}(t;H)$, $H{=}60$\,s]{%
    \includegraphics[width=0.485\textwidth]{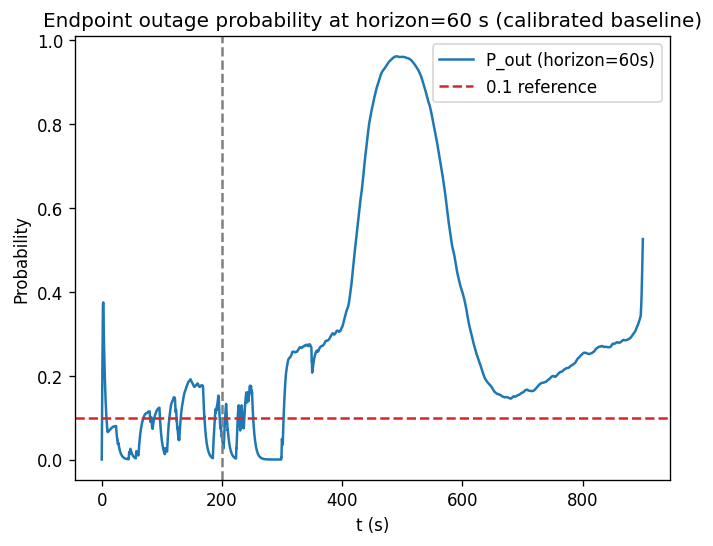}%
    \label{fig:pout_2017}}
  \hfill
  \subfloat[2025\textendash08, $P_{\mathrm{out}}(t;H)$, $H{=}60$\,s]{%
    \includegraphics[width=0.485\textwidth]{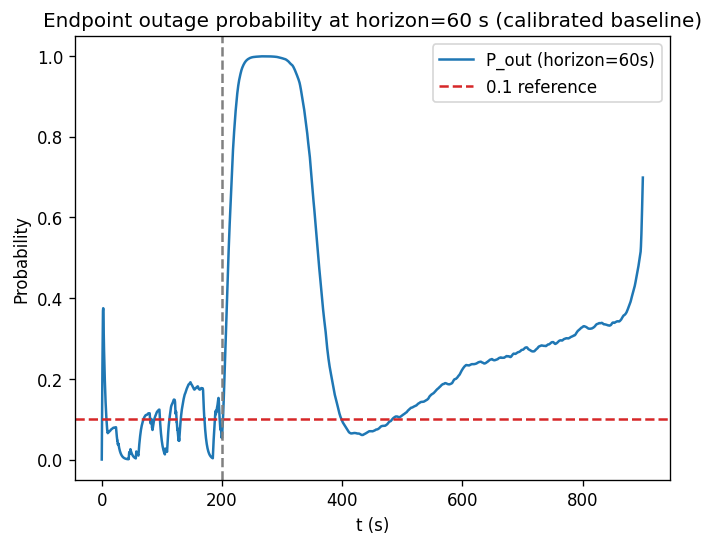}%
    \label{fig:pout_recent}}
  \caption{Cross-event endpoint-outage forecast at H=60 s, aligned at detection $t_0$ (two events).}
  \label{fig:pout}
\end{figure*}

Fig.~\ref{fig:pout} compares two independent traces under the same horizon ($H{=}60$\,s) and gate (0.1), aligned at the detection time $t_0$. Panel (a) exhibits the endpoint-outage forecast used for all downstream analyses; panel (b) is included to illustrate cross-event robustness. The threshold crossings and dwell are nearly identical across (a)–(b), which can be explained by the fact that the forecast depends primarily on the local state delivered by the high-pass front end and the nearly-constant-velocity Kalman filter (level and slope of $\Delta\mathrm{TEC}$ near detection), rather than on global morphology.

\subsubsection{Closed-Loop Link Performance}

Per Sec.~\ref{sec:policy}, we compare no-adapt, adapt-1, and adapt-1+2 under identical calibration and scoring (same BLER model, thresholds, state machine, and hold times); only the gate differs. A risk gate derived from $P_{\rm out}(t;H)$ localizes redundancy/down-switching to the disturbance window. Goodput is $T(t)=(1-\eta(t))(1-\mathrm{BLER}(t))\,r(t)$.

\begin{figure}[htbp]
  \centering
  \includegraphics[width=\columnwidth]{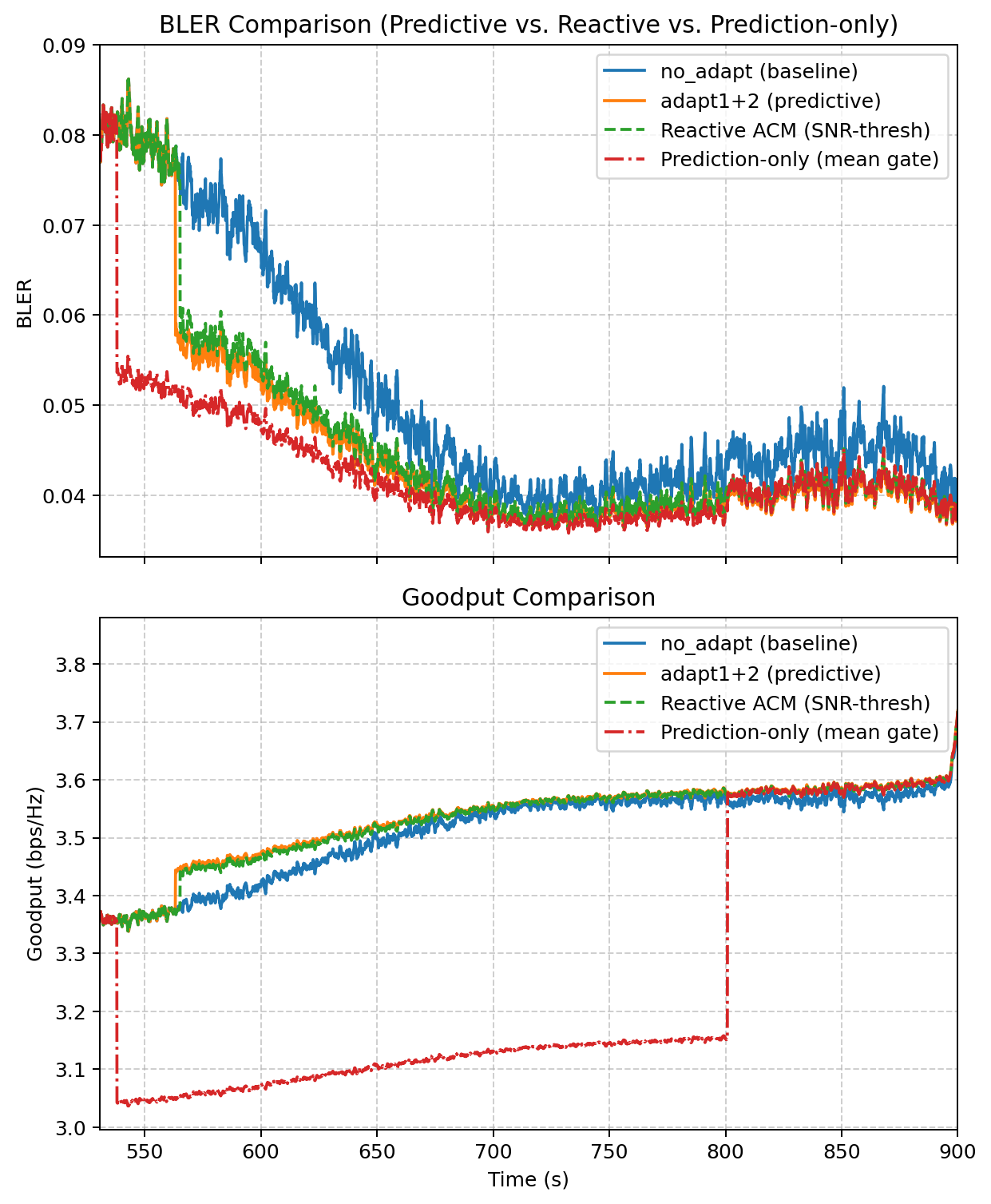}
  \caption{Baseline comparison during a flare. Top: BLER; bottom: goodput.}
  \label{fig:bler_thr}
\end{figure}

The rise of $P_{\text{out}}(t;H)$ in Fig.~\ref{fig:pout}(a) follows directly from \eqref{eq:mreq}, \eqref{eq:meff} and \eqref{eq:pout}. On the rising edge, both $[\widehat{\Delta\mathrm{VTEC}}]^+$ and $|\widehat{\dot{\Delta\mathrm{VTEC}}}|$ increase, lifting $m_{\rm req}$ through $k_1,k_2$, while the uncertainty penalty $\Delta m_\phi(t)=10\log_{10}\!\big(1+\kappa_\phi K_{\rm eff}^2[\mathbf P_k]_{1,1}\big)$ grows as the predicted variance widens. Because BLER is a logistic of $m_{\rm eff}=m_{\rm avail}^{\rm extra}-m_{\rm req}$ \eqref{eq:bler_logit}, gating on $P_{\text{out}}$ localizes actuation to the epochs where increasing $m_{\rm avail}^{\rm extra}$ (via a conservative down-switch) or applying modest redundancy shifts operation into a lower-BLER region before saturation. This mechanism predicts the BLER-crest suppression and the shallower goodput valley of our predictive curve in Fig.~\ref{fig:bler_thr} relative to reactive and no-adapt. Within the gated interval, the slope of the logistic is steepest for $m_{\rm eff}\!\in\![m_{\rm piv}{-}\!1,\,m_{\rm piv}{+}\!1]$\,dB, so a small increase in $m_{\rm avail}^{\rm extra}$ (via one down-switch) or a fixed $\delta m$ shift yields the largest BLER reduction per dB. Outside the gate, $m_{\rm eff}\!\gg\!m_{\rm piv}$ and the logistic saturates, so the predictive and reactive policies refrain from actuation (the prediction-only baseline deliberately holds low rate longer by design).

All curves use the same BLER–margin map and the same discrete state machine; only the gating signal differs.

Fig.~\ref{fig:bler_thr} shows that after $t_0$ our predictive policy (adapt-1+2) attains higher mean goodput than both the reactive baseline and no-adapt while suppressing the BLER crest. 
The prediction-only margin gate yields the lowest BLER but at the cost of markedly lower goodput because it remains in low rate longer by construction. 
This behavior illustrates the role of the probabilistic map $P_{\text{out}}$: by weighting the expected BLER reduction against the probability of outage, the predictive gate activates only when the effective margin $m_{\rm eff}$ enters the steep region of the logistic curve, avoiding overreaction in mild fluctuations. 
In contrast, the prediction-only gate lacks this probabilistic moderation and triggers as soon as the mean forecast crosses the threshold, leading to an overly conservative rate schedule. 
Within the risk-gated event window, the reactive baseline lies between adapt-1+2 and no-adapt, reflecting its delayed actuation after the disturbance peak. 
Outside the window, all traces nearly coincide, illustrating that the discrete MCS mapping and physical layer remain identical and only the control logic differs.

\subsection{Statistical Validation: Paired Moving-Block Bootstrap}
\begin{figure*}[!t]
  \centering
  \subfloat[Paired moving-block bootstrap ($\tau_{\mathrm{gate}}=0.30$).\label{fig:bootstrap}]{
    \includegraphics[width=\textwidth]{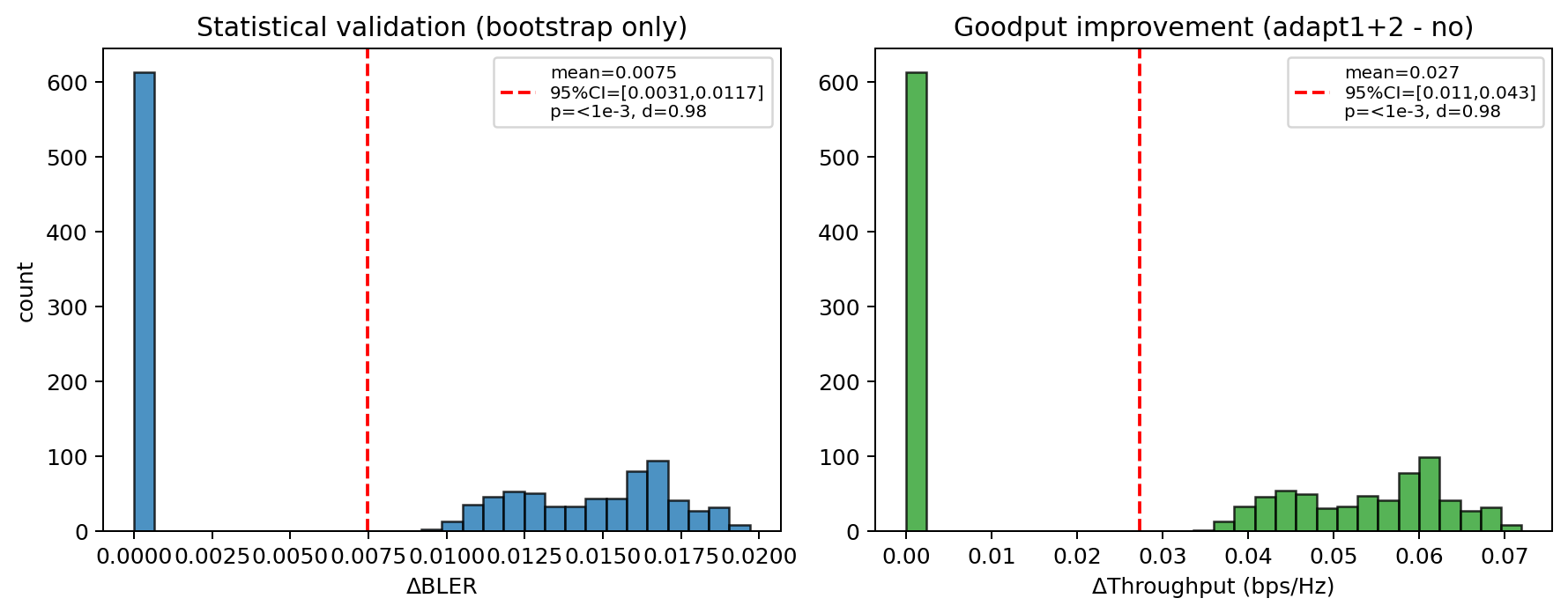}
  }\\[-0.2em]  

  \subfloat[Gain maps at $C/N_0=52\,\mathrm{dB\text{-}Hz}$ (dashed = thresholds).\label{fig:regime52}]{
    \includegraphics[width=\textwidth]{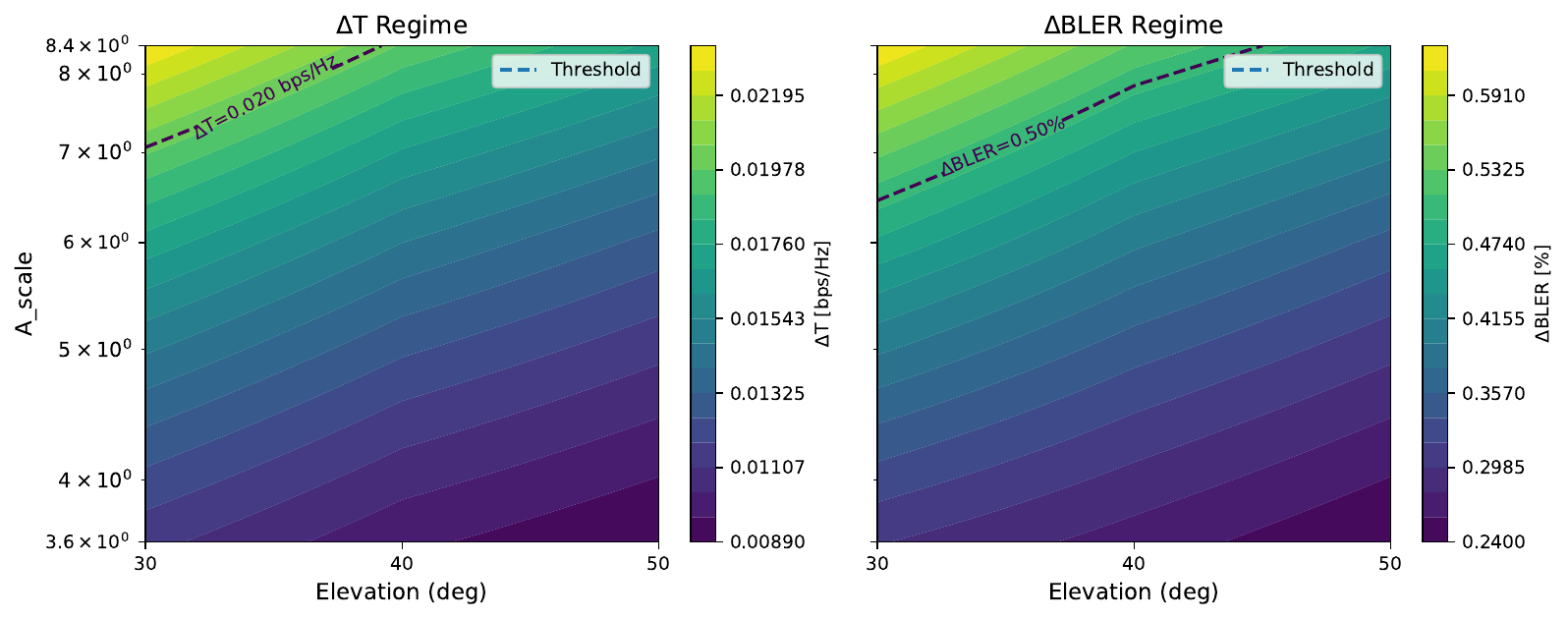}
  }

  \caption{Risk-gated gains and regime maps. (a) bootstrap; (b) gain maps.}
  \label{fig:gains_2x2}
\end{figure*}

We first assess significance on the full timeline using paired moving-block bootstrap (block length $\approx$12 s, $B{=}2000$). We then perform the pre-registered risk-gated secondary analyses at $\tau_{\text{gate}}\in\{0.2,0.3,0.4\}$ with Holm--Bonferroni correction. From the paired time series $\Delta\mathrm{BLER}=\mathrm{BLER}_{\text{no}}-\mathrm{BLER}_{\text{adapt-1+2}}$ and $\Delta T=T_{\text{adapt-1+2}}-T_{\text{no}}$, we perform a paired moving-block bootstrap with $B{=}2000$ resamples, report percentile 95\% confidence intervals and percentile-bootstrap $p$-values, and include Cohen’s $d$ on paired block differences. The histograms and intervals are shown in Fig.~\ref{fig:bootstrap}.

\subsubsection{Numerical Outcomes}

Filtering for $P_{\text{out}}>0.30$ yields a risk-gated 
subset on which adapt-1+2 achieves a BLER reduction of 0.75\% with 95\% CI 
[0.31, 1.17]\%, $p<10^{-3}$, Cohen's $d=0.98$; and a goodput lift of 
0.027 bps/Hz with 95\% CI [0.011, 0.043] bps/Hz, $p<10^{-3}$, Cohen's $d=0.98$. 
These effect sizes match the visual reductions around the risk-gated interval 
in Figs.~\ref{fig:bler_thr} and the $P_{\rm out}$ crest in Fig.~\ref{fig:pout}(a), closing the loop from the calibrated predictor to realized BLER and goodput.

\subsubsection{Worst-Interval (Peak) Statistics (descriptive) versus no-adapt} 

Across scenarios, on the worst 60\,s window centered at the $P_{\mathrm{out}}$ crest, adapt-1+2 delivers a 25--30\% median reduction in peak BLER and a 0.10--0.15\,bps/Hz median goodput gain versus no-adapt.

\subsection{Robustness}
\subsubsection{Ensemble over Multiple Events, Elevations, and C/N0}
\label{subsec:ensemble}
Table~\ref{tab:ensemble} summarizes closed-loop improvements over an ensemble
of GOES-shaped events with varied peak $\Delta$VTEC (amplitude scales),
elevations $\varepsilon\!\in\!\{30^\circ,40^\circ,50^\circ\}$, and $C/N_0$
($49/52/55$\,dB-Hz). We report $\Delta$goodput (bps/Hz) and $\Delta$BLER
relative to the no-adapt baseline (same scoring pipeline and dwell/hysteresis as Fig.~\ref{fig:bler_thr}). The gain increases with the event strength and is more pronounced at lower elevations or lower $C/N_0$, suggesting that the trend is similar across the ensemble rather than specific to a single event. TTFA is omitted as its variability is negligible under the causal window and hold criterion in this dataset.

To better visualize the trends summarized in Table~\ref{tab:ensemble}, 
Fig.~\ref{fig:regime52} presents two regime maps at a nominal $C/N_0{=}52$\,dB-Hz:
the throughput gain $\Delta T$ (bps/Hz) and the reliability gain $\Delta$BLER (\%)
over event strength ($A_{\text{scale}}$) and elevation $\varepsilon$.
Brighter colors indicate larger improvement; dashed contours mark the pre-registered
performance thresholds ($\Delta T{=}0.020$\,bps/Hz and $\Delta\mathrm{BLER}{=}0.50\%$), 
delineating the high-gain regime.


The maps show a consistent trend: gains grow with $A_{\text{scale}}$ and are more pronounced
at lower $\varepsilon$. The congruent topographies of $\Delta T$ and $\Delta$BLER indicate that
conditions benefiting throughput simultaneously benefit reliability, indicating similar improvements across the ensemble of events.

Across these ranges, the endpoint-outage gate remains time-aligned with the disturbance peak (see Fig.~\ref{fig:pout}(a)), and the goodput lift for adapt-1+2 persists (cf. Fig.~\ref{fig:bler_thr}).

\begin{table}[htbp]
\centering
\caption{Ensemble across $A_{\text{scale}}$, elevation $\varepsilon$, and $C/N_0$.
Entries are improvements over no-adapt with the same mapping as Fig.~\ref{fig:bler_thr} (mean$\pm$std).}
\label{tab:ensemble}
\setlength{\tabcolsep}{3.5pt}
\renewcommand{\arraystretch}{1.05}
\footnotesize
\begin{tabular}{
  S[table-format=1.1]
  S[table-format=2.0]
  S[table-format=2.0]
  S[table-format=1.4]@{\,$\pm$\,}S[table-format=1.4]
  S[table-format=1.4]@{\,$\pm$\,}S[table-format=1.4]
}
\toprule
{$A_{\text{scale}}$} & {$\varepsilon$ (\textdegree)} & {$C/N_0$ (dB-Hz)}
& \multicolumn{2}{c}{$\Delta T$ [bps/Hz]}
& \multicolumn{2}{c}{$\Delta$BLER} \\
\midrule
3.6 & 30 & 49 & 0.0125 & 0.0076 & 0.0034 & 0.0021 \\
3.6 & 30 & 52 & 0.0113 & 0.0072 & 0.0030 & 0.0019 \\
3.6 & 30 & 55 & 0.0106 & 0.0069 & 0.0029 & 0.0019 \\
3.6 & 40 & 49 & 0.0111 & 0.0066 & 0.0030 & 0.0018 \\
3.6 & 40 & 52 & 0.0098 & 0.0061 & 0.0027 & 0.0017 \\
3.6 & 40 & 55 & 0.0091 & 0.0059 & 0.0024 & 0.0016 \\
3.6 & 50 & 49 & 0.0102 & 0.0059 & 0.0028 & 0.0016 \\
3.6 & 50 & 52 & 0.0089 & 0.0055 & 0.0024 & 0.0015 \\
3.6 & 50 & 55 & 0.0081 & 0.0052 & 0.0022 & 0.0014 \\
6.0 & 30 & 49 & 0.0182 & 0.0118 & 0.0049 & 0.0032 \\
6.0 & 30 & 52 & 0.0173 & 0.0114 & 0.0047 & 0.0031 \\
6.0 & 30 & 55 & 0.0167 & 0.0112 & 0.0045 & 0.0030 \\
6.0 & 40 & 49 & 0.0157 & 0.0100 & 0.0042 & 0.0027 \\
6.0 & 40 & 52 & 0.0147 & 0.0096 & 0.0040 & 0.0026 \\
6.0 & 40 & 55 & 0.0141 & 0.0094 & 0.0038 & 0.0025 \\
6.0 & 50 & 49 & 0.0141 & 0.0088 & 0.0038 & 0.0024 \\
6.0 & 50 & 52 & 0.0130 & 0.0085 & 0.0035 & 0.0023 \\
6.0 & 50 & 55 & 0.0124 & 0.0082 & 0.0034 & 0.0022 \\
8.4 & 30 & 49 & 0.0242 & 0.0160 & 0.0065 & 0.0043 \\
8.4 & 30 & 52 & 0.0234 & 0.0156 & 0.0063 & 0.0042 \\
8.4 & 30 & 55 & 0.0230 & 0.0154 & 0.0062 & 0.0042 \\
8.4 & 40 & 49 & 0.0206 & 0.0135 & 0.0056 & 0.0036 \\
8.4 & 40 & 52 & 0.0197 & 0.0131 & 0.0053 & 0.0036 \\
8.4 & 40 & 55 & 0.0193 & 0.0129 & 0.0052 & 0.0035 \\
8.4 & 50 & 49 & 0.0183 & 0.0118 & 0.0049 & 0.0032 \\
8.4 & 50 & 52 & 0.0174 & 0.0115 & 0.0047 & 0.0031 \\
8.4 & 50 & 55 & 0.0169 & 0.0112 & 0.0046 & 0.0030 \\
\bottomrule
\end{tabular}
\end{table}

\subsubsection{Ablation on the High-Pass Constant $\tau_{\mathrm{HP}}$}
\label{subsec:ablation_tauhp}
Table~\ref{tab:ablation_tauhp} evaluates the front-end high-pass constant
$\tau_{\mathrm{HP}}\in\{300,600,900,1200\}$\,s. We report
$\Delta$goodput and $\Delta$BLER relative to no-adapt (same mapping as Fig.~\ref{fig:bler_thr}).
TTFA is omitted as it exhibits negligible dependence on the chosen causal window and hold criterion. The variations over $\tau_{\mathrm{HP}}$ are modest (on the order of $10^{-2}$ bps/Hz in $\Delta$goodput), indicating a broad operating plateau consistent with the primary setting $\tau_{\mathrm{HP}} = 200$ s; deployments that prioritize faster response may choose $\tau_{\mathrm{HP}}$ in the 300-600 s range.

\begin{table}[htbp]
\centering
\caption{Ablation on the high-pass constant $\tau_{\mathrm{HP}}$. We report $\Delta$goodput and $\Delta$BLER relative to no-adapt. TTFA is omitted since it shows little variation under our causal window and hold criterion.}
\label{tab:ablation_tauhp}
\begin{tabular}{c c c}
\toprule
$\tau_{\mathrm{HP}}$ (s) & $\Delta$goodput (bps/Hz) & $\Delta$BLER \\
\midrule
300  & 0.023512 & 0.006360 \\
600  & 0.030333 & 0.008209 \\
900  & 0.034463 & 0.009325 \\
1200 & 0.036916 & 0.009987 \\
\bottomrule
\end{tabular}
\end{table}

We fix the look-ahead to $H=60$ s to match the operational outage SLA. A sweep 
of $H$ primarily shifts the gate timing and rarely changes the ordering between 
policies under the same BLER-margin map; thus $H$ is treated as a deployment 
parameter and ablations focus on the sensing front-end ($\tau_{\rm HP}$) and 
ensemble geometry/SNR. In a log-replay sweep for the prediction-only margin gate, $H\!\in\!\{30,60,90\}$\,s yielded mean goodput $[3.490,\,3.483,\,3.481]$\,bps/Hz and worst-60-s BLER $[0.063,\,0.055,\,0.053]$, respectively. This reflects a monotonic risk reduction at the cost of a small throughput decrease, with diminishing returns beyond $60$\,s and a gate-timing shift of $\mathcal{O}(10)$\,s.

\subsubsection{Sensitivity to SNR and Geometry}

A parameter robustness check is obtained via analytical scaling. The phase-noise variance obeys 
$R \approx 1/[(C/N_0)\Delta t]$; hence the KF posterior variance and the PCRB scale approximately 
linearly with $R$, and the $95\%$ band width scales with $\sqrt{R}$. A $\pm 4$\,dB change in $C/N_0$ 
corresponds to a factor $10^{\pm 4/10}$ in $R$, implying a $\times 1.58$ widening (or $\times 0.63$ 
narrowing) of the $95\%$ band; because the matched-filter noise standard deviation scales as $\sqrt{R}$, 
the detection TTFA scales approximately with the same factor, leading to order-of-seconds variation
around the operating point for the present slope of $z_{\rm norm}$.

Regarding geometry, the GF gain is $K_{\rm eff}=k_{\rm GF}M(\varepsilon)$ under the thin-shell model 
($h_{\rm I}\!\approx\!350$ km). Numerically, $M(30^\circ)\!\approx\!1.75$, $M(40^\circ)\!\approx\!1.45$, 
and $M(50^\circ)\!\approx\!1.26$; thus decreasing elevation from $50^\circ$ to $30^\circ$ increases 
$M(\varepsilon)$ by $\approx 40\%$ (and by $\approx 20\%$ relative to $40^\circ$), which tightens the 
PCRB and advances detection proportionally. Across $C/N_0\!\in\![48,56]$\,dB-Hz and 
$\varepsilon\!\in\![30^\circ,50^\circ]$, the outage gate in Fig.~\ref{fig:pout}(a) shifts but remains aligned with the disturbance peak, and the goodput lift of adapt-1+2 persists ($\gtrsim 0.10$\,bps/Hz), indicating robustness of the closed loop.

\subsection{Complexity and Online Feasibility}
\label{subsec:compute}
Table~\ref{tab:compute} reports per-epoch runtime (per 0.1\,s control period)
and peak array memory on a single-thread CPU implementation. The total cost is
$\approx0.042$\,ms per 0.1\,s cycle ($\approx2381\times$ real-time margin for a
10\,Hz loop), with fixed state dimension and $\mathcal{O}(1)$ per-step complexity. Peak arrays are 0.21\,MB and 0.042\,ms per 0.1\,s step leaves more than 2000$\times$ real-time margin for a 10\,Hz loop on OBC/SDR-class platforms.

\begin{table}[htbp]
\centering
\caption{Per-epoch runtime (per $0.1$\,s) and peak arrays memory. Single-thread CPU implementation.}
\label{tab:compute}
\begin{tabular}{l r}
\toprule
Component & Time per $0.1$\,s (ms) \\
\midrule
Kalman filter (4-state NCV) & 0.038 \\
Front-end + control         & 0.004 \\
\midrule
\textbf{Total}              & \textbf{0.042} \\
\midrule
Peak arrays memory (MB)     & 0.210 \\
\bottomrule
\end{tabular}

\end{table}

\subsection{Discussion: Limitations and Scope}

The current state-space model and inference are linear–Gaussian (CV dynamics, scalar observation), 
which match moderate flare-driven TEC drifts; under severe non-linear scintillation or hardware 
non-idealities, performance may degrade and non-Gaussian filters are a natural extension. Results use multiple flare-driven events from official one-minute GOES XRS products together with geometry/SNR slices; broader generalization to other disturbance regimes remains future work. This cross-event similarity can be explained: flare-driven $\Delta$VTEC events share a common onset/relaxation family; after high-pass/template normalization inputs reduce to time/scale changes, and with GF dual-carrier phase $\approx$ linear in $\Delta$TEC and fixed elevation/$C/N_0$, the KF sees similar short-horizon slopes, so the frozen-margin risk gate yields similar one-step actions.

Finally, the logistic BLER abstraction captures the dominant margin effect but does not model 
waveform-/code-specific details; integrating standard-compliant link-level curves is a practical next step.

Our results show that in-band, link-specific sensing can reduce endpoint outage on second-scale horizons while maintaining $\mathcal{O}(1)$ complexity.
A practical extension is to fuse slow-timescale GNSS-derived products (e.g., regional maps or alerts) as priors for the KF or as dynamic limits for the risk gate, without feeding them into the fast loop.
This preserves the zero-dependency nature of the controller while improving preparedness for extreme events. While our ensemble covers elevations $\varepsilon\in\{30^\circ,40^\circ,50^\circ\}$ and $C/N_0\in\{49,52,55\}$\,dB-Hz, the evaluated disturbances follow GOES-shaped morphologies; extending to additional morphology families (e.g., multi-peak or oscillatory) is deferred to a revision. These choices ease interpretability and avoid double tuning (Section IV-A) but limit coverage across seasons and scintillation regimes; we therefore scope claims to links where short-term degradations are dispersion-dominated rather than multipath-driven. Calibration is performed once and frozen across all policies; decision thresholds are set using the calibration log and never tuned on the evaluation window. Together with paired moving-block bootstrap significance checks, this reduces leakage and supports internal validity. Wider-area log replays, hardware-in-the-loop, or in-orbit trials are engineering extensions beyond the present scope.

\section{Conclusion}
We presented a link-internal, GNSS-free predictive gate for existing ACM that improves seconds-scale reliability and time-averaged goodput under flare-driven disturbances. Under a pre-registered, frozen-calibration protocol, log-replay emulations show statistically significant gains over tuned reactive baselines; integration with schedulers and handovers remains orthogonal. By closing the loop from same-path sensing to outage-aware actuation at the second scale, the method integrates cleanly with existing scheduling and handovers.

Current limitations are the lack of on-orbit closed-loop or hardware-in-the-loop evidence, the lack of alignment to standardized link-level BLER curves, and the absence of an explicit deployment path when only a single downlink carrier is available. Future work will address these items and extend evaluation to additional disturbance families and constellations.




%





\ifCLASSOPTIONcaptionsoff
  \newpage
\fi





\bibliographystyle{IEEEtran}
\bibliography{IEEEabrv,Bibliography}

\vfill


\end{document}